%% file: ms.tex
\documentclass[letterpaper,twocolumn,10pt]{article}
\usepackage{usenix2019_v3}

\usepackage{breakurl}
\usepackage{tikz}
\usepackage{amsmath}
\usepackage{comment}
\usepackage{cite}
\usepackage{amsmath,amssymb,amsfonts}
\usepackage{algorithmic}
\usepackage{graphicx}
\usepackage{textcomp}
\usepackage{xcolor}
\usepackage{xspace}
\usepackage{booktabs}
\usepackage{multirow}
\usepackage{soul} 
\usepackage[htt]{hyphenat} 
\usepackage{url} 
\usepackage{makecell}
\usepackage{subcaption} 
\usepackage{todonotes} 
\usepackage{graphicx} 
\usepackage{filecontents}
\usepackage{pifont}
\usepackage[shortlabels]{enumitem}
\usepackage[toc,title,page]{appendix}
\usepackage{subcaption} 
\usepackage[subtle]{savetrees}

\usepackage[noindentafter]{titlesecfix}
\titlespacing\section{0pt}{12pt plus 4pt minus 2pt}{2pt plus 2pt minus 2pt}
\titlespacing\subsection{0pt}{12pt plus 4pt minus 2pt}{1pt plus 2pt minus 2pt}
\titlespacing\subsubsection{0pt}{12pt plus 4pt minus 2pt}{0pt plus 2pt minus 2pt}

\def\BibTeX{{\rm B\kern-.05em{\sc i\kern-.025em b}\kern-.08em
    T\kern-.1667em\lower.7ex\hbox{E}\kern-.125emX}}

\newcommand{\sysname}{{{\fontfamily{lmss}\selectfont PushAdMiner}}\xspace}
\newcommand{\karthika}[1]{\textcolor{purple}{[(Karthika): #1]}}
\newcommand{\phani}[1]{\textcolor{brown}{[(Phani): #1]}}

\newcommand{\KYU}{\textcolor{blue}}

\begin{document}
 
\date{}

\title{
Measuring Abuse in Web Push Advertising 
}

\author{
{\rm Karthika Subramani}\\
University of Georgia
\and
{\rm Xingzi Yuan}\\
University of Georgia
\and
{\rm Omid Setayeshfar}\\
University of Georgia
\and
{\rm Phani Vadrevu}\\
University of New Orleans
\and
{\rm Kyu Hyung Lee}\\
University of Georgia
\and
{\rm Roberto Perdisci}\\
University of Georgia and Georgia Tech
} 

\maketitle
\titlespacing{\section}{0pc}{1pc}{0.3pc}
\titlespacing{\subsection}{0pc}{0.8pc}{0.3pc}
\titlespacing{\subsubsection}{0pc}{0.6pc}{0.3pc}
\input{abstract.tex}

\input{intro.tex}

\input{background.tex}

\input{system_overview.tex}

\input{data_collection.tex}

\input{data_analysis.tex}
\input{evaluation.tex}

\input{related.tex}
\input{discussion.tex}

\input{conclusion.tex}

\bibliographystyle{unsrt}
\bibliography{references}
\input{appendix.tex}

\end{document}

%% file: abstract.tex
\begin{abstract}

The rapid growth of online advertising has fueled the growth of ad-blocking
software, such as new ad-blocking and privacy-oriented browsers or browser
extensions. In response, both ad publishers and ad networks are constantly
trying to pursue new strategies to keep up their revenues. To this end, ad
networks have started to leverage the {\em Web Push} technology enabled by
modern web browsers.

As web push notifications (WPNs) are relatively new, their role in ad delivery
has not been yet studied in depth. Furthermore, it is unclear to what extent WPN
ads are being abused for {\em malvertising} (i.e., to deliver malicious ads). In
this paper, we aim to fill this gap. Specifically, we propose a system called
\sysname that is dedicated to (1) {\em automatically} registering for and
collecting a large number of web-based push notifications from publisher
websites, (2) finding WPN-based ads among these notifications, and (3)
discovering malicious WPN-based ad campaigns.

Using \sysname, we collected and analyzed 21,541 WPN messages by visiting
thousands of different websites. Among these, our system identified 572 WPN ad
campaigns, for a total of 5,143 WPN-based ads that were pushed by a variety of
ad networks. Furthermore, we found that 51\% of all WPN ads we collected are
malicious, and that traditional ad-blockers and malicious URL filters are
remarkably ineffective against WPN-based malicious ads, leaving a significant
abuse vector unchecked.

\end{abstract}

%% file: intro.tex
\section{Introduction}
\label{sec:intro}

In the past few years, the rapid growth of online advertising has fueled the
growth of ad-blocking software, such as new ad-blocking and privacy-oriented
browsers (e.g., Brave~\cite{brave}) or browser extensions (e.g.,
AdBlockPlus~\cite{adblockplus}). In response, both ad publishers
and ad networks are constantly trying to pursue new strategies to keep up their
revenues. To this end, ad networks have started to leverage the {\em Web Push}
technology enabled by modern web browsers~\cite{WebPushNotifications}. Until
relatively recently, push notifications were mostly limited to native apps on
mobile platforms, and web-based applications were unable to connect to their
users out of active browsing sessions. However, now Web Push allows for web
applications to send out Web Push Notifications (WPN) at any time to re-engage
their users, even when the browser tab in which the web application was running
is closed (the browser itself needs to be running, but does not need to be in
the foreground for a WPN to be delivered to the user). Furthermore, unlike push
notifications from native mobile apps, WPNs allow for notifications to be
displayed on both desktop and mobile devices. Thus, they serve as a single tool
with support to reach users on multiple platforms.

Although WPNs were initially designed for websites to deliver simple messages
(e.g., news, weather alerts, etc.), they have become an effective way to also
serve online ads, and can therefore be abused to also deliver malicious ads. In
particular, the use of WPNs for ad delivery has some unique advantages. First,
unlike traditional online ads (banner ads, pop-up ads or pop-under ads),
advertisers do not have to wait for users to reach the web page that publishes
the ad. Instead, advertisers can send out notifications that can allure users to
their targeted content. Secondly, thanks to years of experience with native
mobile app notifications, users have been trained to compulsively interact with
push notification messages (at least on mobile devices). WPN-based ads may also
be less prone to \emph{ad blindness}~\cite{ad_blindness}, compared to
traditional web ad delivery mechanisms such as page banners. Furthermore,
ad-blocking software are not currently effective at blocking WPN-based ads (see
Section~\ref{sec:ad_blockers}), in part because browser extensions are not
allowed to interfere with the \emph{Service Workers} code through which WPNs are
delivered~\cite{master_web_puppets}. For these reasons, some ad networks are
focusing their business specifically around WPN ads (e.g.,
RichPush~\cite{rishpush}).

As WPNs are relatively new, their role in ad delivery has not been yet studied
in depth. Furthermore, it is unclear to what extent WPN ads are being abused for
{\em malvertising} (i.e., to deliver malicious ads). In this paper, we aim to
fill this gap. Specifically, we propose a system called \sysname that is
dedicated to (1) {\em automatically} registering for and collecting a large
number of web-based push notifications from publisher websites, (2) finding
WPN-based ads among these notifications, and (3) discovering malicious WPN-based
ad campaigns. To build \sysname, we significantly extend the Chromium browser
instrumentations developed by~\cite{JSgraph} and~\cite{Secma}, which have been
open-sourced by the respective authors. Specifically, neither~\cite{JSgraph}
or~\cite{Secma} is able to track the activities of Service Workers in detail.
Therefore, we implement our own set of browser instrumentations that allows us
to track WPNs in all their aspects, from registration to notification delivery,
on both desktop and mobile devices. We then build a custom WPN crawler around
our instrumented browser to automatically receive, track, and interact with
generic WPNs, including collecting malicious WPN ads and their respective
malicious landing pages. Finally, we develop a data mining pipeline to analyze
the collected WPNs and discover malicious WPN-based campaigns.

To the best of our knowledge, ours is the first systematic study that focuses on
automatically collecting and analyzing WPN-based ads and on discovering
malicious ad campaigns delivered via WPNs. In contrast, previous work focused on
other security-related aspects of Service Workers and Push Notifications, such
as building stealthy botnets~\cite{master_web_puppets}, or social engineering
attacks that attempt to force users into subscribing to push
notifications~\cite{Secma} but without studying the resulting push messages. Lee
at al.~\cite{pride} study Progressive Web Apps. They collect Service Worker
scripts from top-ranked website homepages and analyze their push notifications.
Their work studies potential security vulnerabilities related to Service
Workers, App Cache, and discusses how push notifications may be abused to
launch phishing attacks, without measuring how prevalent these attacks are in
the wild. Our work is different, in that we aim to {\em automatically} collect
and analyze WPN-based ads, to discover WPN ad campaigns, and to measure the
prevalence of malicious WPN-based ad campaigns in the wild.

In summary, we make the following contributions:
\begin{itemize}
\setlength\itemsep{0em}
\item We present \sysname, a system that enables the automated collection and
analysis of online ads delivered via web push notifications (WPNs) on both
desktop and mobile devices.
\item To track WPNs, we significantly extend a Chromium-based instrumented
browser developed in~\cite{JSgraph, Secma} to allow for a detailed analysis of
Service Workers, which are at the basis of WPN deliveries. Furthermore, we build
a custom WPN crawler around our instrumented browser to collect and
automatically interact with WPNs.
\item  Using \sysname, we collected and analyzed 21,541 WPN messages by visiting
thousands of different websites. Among these, our system identified 572 WPN ad
campaigns, for a total of 5,143 WPN-based ads that were pushed by a variety of
ad networks. Furthermore, we found that 51\% of all WPN ads we collected are
malicious, and that traditional ad-blockers and malicious URL filters are
remarkably ineffective against WPN-based malicious ads, leaving a significant
abuse vector unchecked.
\end{itemize}

%% file: background.tex
\section{Motivating Example and Background}
\label{sec:background}

\begin{figure*}[ht]
\setlength\belowcaptionskip{-0.7\baselineskip}
\setlength\abovecaptionskip{0.4\baselineskip}
\centering
\includegraphics[width=\textwidth]{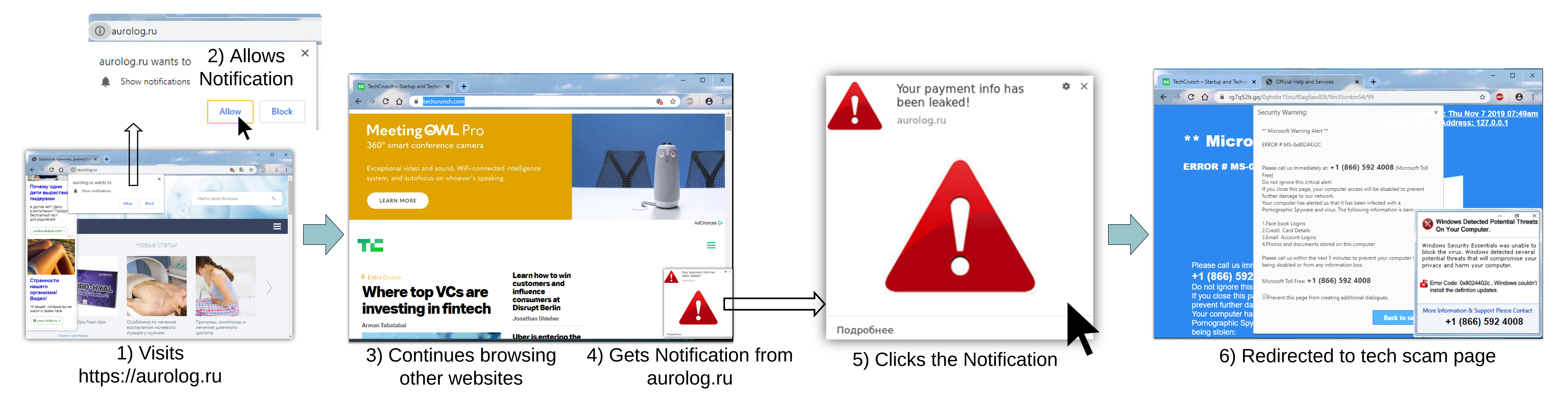}
\caption{Example of malicious advertisement served through web push notifications}
\label{tech_scam}
\end{figure*}

In this section, we provide an example of WPN-based malicious ad, and then
briefly explain the concepts and technologies behind web notification services.

\subsection{Motivating Example}
\label{sec:motivating_example}

Figure~\ref{tech_scam} provides an example of malicious WPN-based ad. During the
preliminary stages of our research, we stumbled upon a website on
\url{aurolog[.]ru}. When visiting the main page, the site requested permission
to send us notifications. We granted permission by pressing the {\em Allow}
button on the browser dialog box, and subsequently received a WPN ad with the
following alert message: ``Your payment info has been leaked'' (see
Figure~\ref{tech_scam}). After clicking on the notification, we were redirected
to a {\em tech support scam}~\cite{dial_one_scam}. To our surprise, the landing
URL was neither blacklisted by Google Safe
Browsing\cite{gsb} nor detected as malicious by
any of the web page scanners on Virus Total\cite{virus_total}. This
example confirmed our suspicion that WPNs may be abused for malvertising, and
sparked our investigation to determine whether such cases of malicious WPN-based
ads could be automatically collected and analyzed.

\subsection{Technical Background}

Recent changes in HTML5 have introduced new web features, such as
\emph{Service Workers}\cite{service_workers}, \emph{Push
Notifications}\cite{notifications_api} and \emph{AppCache}\cite{app_cache}.
Websites that adopt these technologies are called Progressive Web Apps
(PWAs). Throughout this paper, we refer to push notifications sent by PWAs using
a browser as Web Push Notifications (WPN), to distinguish them from push
notifications sent by native apps on mobile devices.

\noindent 
\textbf{Service Workers and Push Notifications:}
A {\em Service Worker} is an event-driven script executed by the browser in the
background, separately from the main browser thread and independently of the web
application from which it was initially registered and that it controls. In
practice, a Service Worker comes in the form of a JavaScript file that is
registered against the origin and path of the web page to which it is associated
(only HTTPS origins are allowed to register a Service Worker). 
In effect, Service Workers can be viewed as ``a programmable network proxy that
lets you control how network requests from your page are
handled''\cite{service_worker_intro}.

Service Workers can use the {\em Push
API}\cite{push_api}
to receive messages from a server, even while the associated web application is
not running. It is worth noting that a single web app is allowed to register
multiple Service Workers.
Service Workers can also use the {\em Notifications API}\cite{notifications_api}
to display system notifications to the user. A prerequisite is that the web
application must first request permission to display notifications to the user
(only allowed for HTTPS origins). If the user accepts (i.e., clicks on ``Allow''
instead of ``Block'' on the notification request popup) to receive notifications
from the web application's origin, this permission persists across browser
restarts, and until the user explicitly revokes the permission via browser
settings/preferences (notice that non-expert users may find it difficult to
understand, find, and disable notifications in the browser's settings).

Web notification messages have a number of customizable parameters, such as
title, body, target URL, icon image, display image and action buttons. The user
can interact with a notification by either clicking on it, closing it or
performing any custom actions displayed in the notification message. Service
Workers can listen to such user events and take action according to the input.
This includes loading the notification's target URL on a separate tab, following
a user's click on the notification box. Therefore, push notifications can also
be conveniently used to deliver textual and graphical ads, and the user can be
redirected to the advertised product page after clicking on the ad. A simplified
illustration of how ads are served via WPNs is shown in
Figure~\ref{notification_system}(in Section~\ref{sec:data_collection}).

\noindent 
\textbf{Firebase Cloud Messaging (FCM):} 
FCM is a cross-platform messaging solution for Push Notifications. It can serve
as a central authority that mediates the communication between the ad server and
the Service Worker. Upon initial registration, FCM creates a unique registration
ID per user and per Service Worker, which is sent along with an endpoint
URL\cite{push_notifications_intro} to the ad server.
For further details, we refer the reader to FCM online
documentation~\cite{Firebase}.

%% file: system_overview.tex
\begin{figure*}[t]
  \centering
  \setlength\belowcaptionskip{-0.7\baselineskip}
  \includegraphics[scale=0.95]{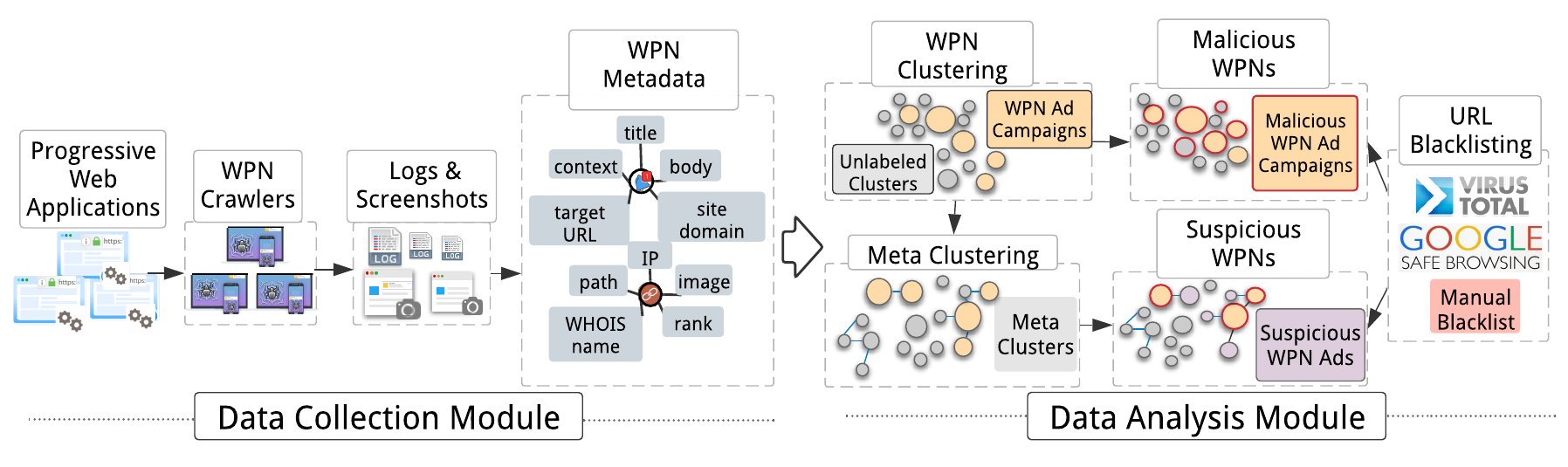}
  \addtolength{\belowcaptionskip}{-5pt}
  \caption{PushAdMiner System Overview}
  \label{fig:sys_overview}
\end{figure*}

\section{System Overview}
\label{sec:sys_overview}

In this section, we provide an overview of how \sysname works, leaving a
detailed description of the main system's components to
Sections~\ref{sec:data_collection} and~\ref{sec:data_analysis}. A high-level
representation of the system is provided in Figure~\ref{fig:sys_overview}.

\sysname consists of three main components: (i) an instrumented browser to
collect fine-grained information about Service Workers and WPNs; (ii) a custom
crawler that automatically visits sites and interacts with the browser,
including granting notification permissions and interacting with WPNs
(Section~\ref{sec:data_collection}); and (iii) a data analysis component aimed
at identifying WPN-based ad campaigns and labeling likely malicious ones
(Section~\ref{sec:data_analysis}).

While a number of browser automation and crawling systems have been proposed,
including Selenium~\cite{Selenium}, Puppeteer~\cite{Puppeteer}, and
others~\cite{JSgraph, Secma, VV8}, currently they do not fully support the automatic
user interactions with WPNs and collection of all details about Service Workers needed
for our study. 
We therefore built an instrumented browser
based on Google Chromium, by significantly extending existing open-source
Chromium instrumentations~\cite{JSgraph, Secma}. In addition, we leveraged
Puppeteer~\cite{Puppeteer} for browser automation and event logging, and wrote
custom scripts to record Service Workers registrations and network requests. We
plan to release the source code of our system and the WPN messages dataset collected for
this study.

Figure~\ref{fig:sys_overview} presents an overview of how \sysname collects
information about WPNs, and how the browser logs are analyzed to identify ad
campaigns in general and discover malicious ones among them. First, we start
with a set of {\em seed} ad networks that are known to specialize in ad delivery
via WPNs. 
Then, we rely on the \url{publicwww.com} engine to perform a ``reverse search''
and find web pages that host Service Workers code distributed by those ad
networks. This process yields a large set of web pages for crawling. After
visiting these pages using our instrumented browser, we automatically determine
which pages request permission to send us notifications. We retain these pages
for further analysis, and discard the others.

For the web pages that ask for notification permissions, we log details about
the responsible Service Worker code, automatically grant permission (via browser
code instrumentation), and then collect the notifications that are later pushed
to our instrumented browser. When a notification is displayed by the browser, we
record fine grained details about the notification message itself (including
message text and icons), automatically simulate a user click on the notification
box (via browser code instrumentation), and track all events resulting from the
click. If the click results in a new page being open, we record detailed
information about the related network requests, including all browser
redirections, as well as detailed logs and a screenshot of each new page the
browser visits, including the landing page (i.e., the final web page reached due
to the click).

Finally, we extract relevant information from the detailed logs of our
instrumented browser, and apply a clustering strategy to find notifications that
are similar to each other, which allows us to identify WPN-based ad campaigns.
We then leverage URL blacklists to find WPN ad campaigns that are likely
malicious (e.g., because one or more landing pages are known to be malicious).
This allows us to prioritize the most suspicious WPN ad campaigns for further
manual analysis and labeling.

Note that in this paper we do not focus on building a malicious
WPN ad campaign detector, such as using statistical features or machine
learning classifiers. Rather, our focus is on discovering, collecting, and
analyzing WPN ad campaigns in general, and on measuring the prevalence of both benign and malicious campaigns. As we will show in
Section~\ref{sec:results}, URL blacklists tend to miss a significant number of
malicious URLs that we determine to be related to malicious ad campaigns. The
analysis we present in this paper could therefore be used as a starting point
for developing an automated malicious WPN ad campaign detector. We leave this
latter task to future work.

\input{ethical.tex}

%% file: ethical.tex
\subsection{Ethical Considerations} 
To track WPN-based ads and label malicious ones, it is necessary to collect
information about the landing page that an ad eventually redirects to. For
instance, for most malicious ads the attack is effectively realized only once
the user reaches the landing page, especially in case of social engineering and
phishing attacks. As we do not know in advance what landing page will be reached
by clicking on a WPN message, and whether a WPN ad is malicious or not, our
system will likely click on both legitimate and malicious ads. In turn, this may
cause legitimate advertisers to incur a small cost for our clicks, as they will
likely have to pay a third-party publishing web page and ad network for their
services (notice that we obviously receive no monetary gain whatsoever during
this process). This is common to other similar studies, such as
\cite{Secma,RafiqueGJHN16}, and we therefore address the ethical considerations
for our study following previous work.

To make sure we do not have a significant negative impact on legitimate
third-parties, we estimated the cost incurred by these advertisers due to ad
clicks performed by our system, and found that our system has minimal impact on
advertisers. Specifically, among the WPN ads we identified, we consider
legitimate ads to be those whose landing pages are not labeled as malicious by
Virus Total's URL classification services. Then, we estimate the {\em cost per
landing domain} based on the number of ads we clicked on that lead to a specific
domain, using the {\em Cost Per Mille (CPM)}~\cite{CPM} for push notification
ads according to iZooto~\cite{izooto_report}. The maximum cost per landing
domain throughout our entire study was USD 1.12\$ (due to landing on the same
domain 444 times), which we calculated using the standard CPM of USD 2.54\$. On
average, we visited each landing domain 18 times, which corresponds to an
average cost of USD 0.04\$ per landing domain (i.e., per advertiser).
Considering these low values, we believe the impact of our system on advertisers
is not significant, and is on par with previous
work~\cite{Secma,RafiqueGJHN16}.

%% file: data_collection.tex
\section{Data Collection Module}
\label{sec:data_collection}

\begin{figure}[ht]
\vspace{-4mm}
\setlength\belowcaptionskip{-0.7\baselineskip}
\setlength\abovecaptionskip{0.4\baselineskip}
\includegraphics[width=\columnwidth]{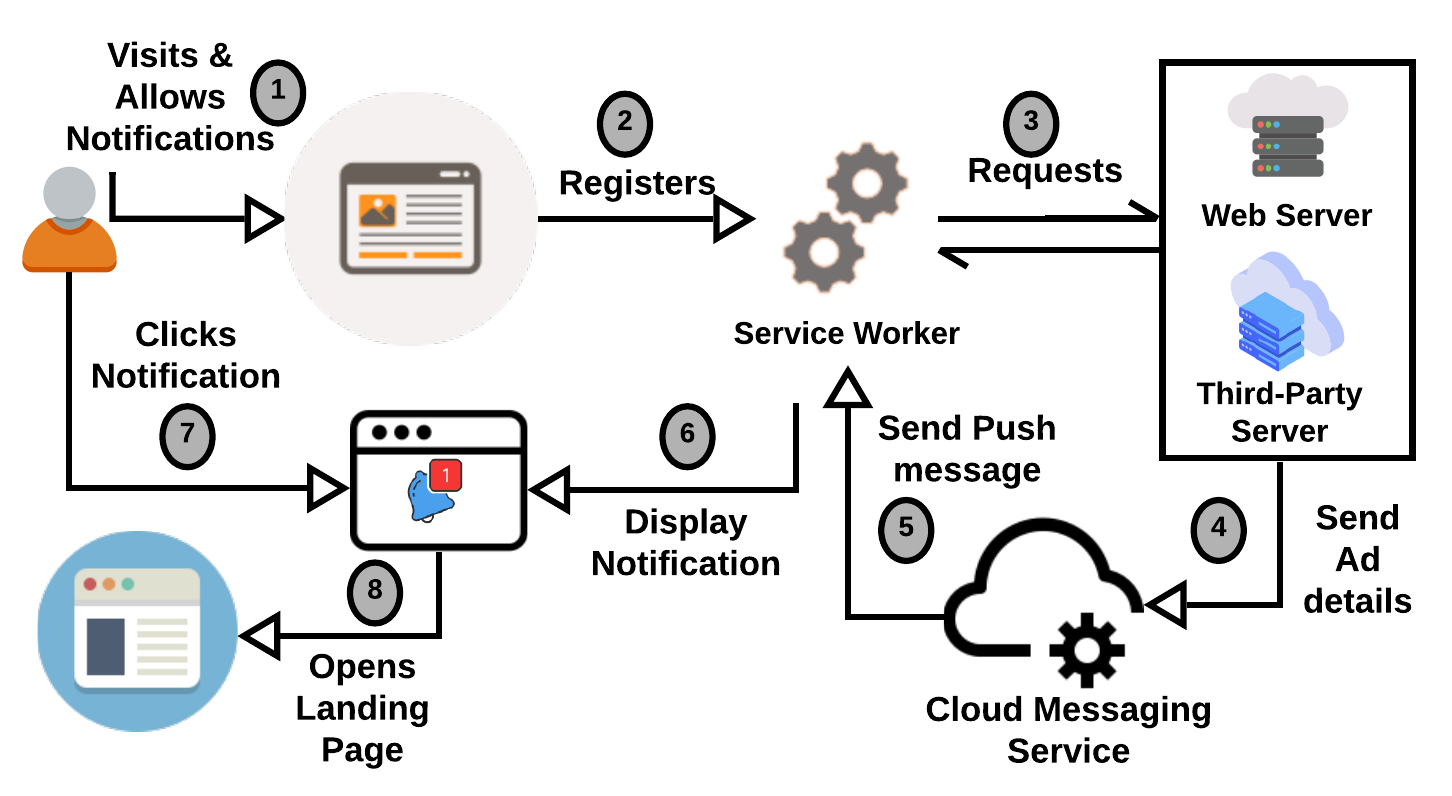}
\caption{Steps involved in Serving Ads via WPNs}
\label{notification_system}
\end{figure}

In this section, we describe in detail how \sysname's data collection modules
work. The steps referred to in the following sections follow the numbering given
in Figure~\ref{notification_system}.

\subsection{Desktop Environment}
To discover WPN ad campaigns, we first need to collect WPN messages. To this
end, we build a crawler consisting of an instrumented browser and browser
automation scripts. As our crawler encounters a website that asks for permission
to send push notifications, our goal is to automatically allow the permission
request, so that we can collect notification from that origin.
To this end, we instrumented our browser as follows: we introduce changes to the
\texttt{RequestPermission} and \texttt{PermissionDecided} methods under the
\texttt{PermissionContextBase} class in Chromium's C++ code base, to log all
details about the permission request (e.g., the origin requesting it) and to
automatically grant permissions.

\vspace{2pt}
\noindent
\textbf{Service WorkerRegistration (step 2): }
To record when a Service Worker is registered by a web page, we use a custom
script for Puppeteer~\cite{Puppeteer}. Specifically, we listen to
\texttt{serviceworkercreated} events and log information such as details about
the page that registered the Service Worker, and the URL from which the Service
Worker code was retrieved. Once the Service Worker has been registered, it can
subscribe for push notifications. During this subscription process, a server API
key is passed to the browser, to identify the server that will be responsible
for sending push notification via the FCM (see Section~\ref{sec:background}). We
log the API key information by listening for \texttt{PushManager.subscribe}
events.

\vspace{3pt}
\noindent
\textbf{Network Requests (step 3): }
When the browser receives a push message from the FCM service, it will alert the
Service Worker to which the message is destined. The correct Service Worker is
identified via a unique ID included in the push message. As part of handling the
push message and related notification, the Service Worker may issue one or more
network requests directly to an ad network server or other third-party servers.
For instance, the Service Worker may contact an ad network server to determine
the landing URL associated to a given push message. Also, after a pushed
notification receives a user click, the Service Worker may send a network
requests to an ad server to notify it of the user click, or to facilitate other
tracking related activities. We capture and log such network requests issued by
Service Workers by leveraging a custom Puppeteer~\cite{Puppeteer} script that
listens to Service Worker \texttt{request} and \texttt{response} events.
Specifically, for every Service Worker's network request we record information
such as the requested URL, the data sent/received, possible network
redirections, and the related response content.

\vspace{3pt}
\noindent
\textbf{Notifications (steps 5-6): }
Notifications are displayed by invoking \texttt{showNotification} under
\texttt{ServiceWorkerRegistrationNotifications} (in Chromium's C++ code). We
therefore add an instrumentation hook to record calls to that method.
Specifically, we log the URL of the Service Worker that called for showing the
notification, the notification title, body, icon URL, and target URL (if
present).

\vspace{3pt}
\noindent
\textbf{Notification Clicks and Navigations (steps 7-8):}
Depending on the type of notification, clicking on it can navigate the browser
to a new web page (e.g., on a separate tab). For WPN-based ads, clicking on the
notification box typically takes the user to the page advertised in the WPN ad
(i.e., the ad's landing page). Collecting such pages is especially important for
studying malicious WPN ads, because the landing page often implements a critical
component of the attack. For instance, Figure~\ref{tech_scam} shows that
the user reaches the actual {\em tech support scam} page after clicking on the
notification. The landing page advertises the scam phone number, without which
the attack could not be monetized.

To automate the process of collecting the landing page associated to push
notification messages, we need to simulate a user click. Unfortunately, browser
automation frameworks such as Puppeteer~\cite{Puppeteer} and
Selenium~\cite{Selenium} do not allow for interacting with WPNs. Therefore, we
again had to build our own custom browser instrumentation. Specifically, we
found that Chromium uses the \texttt{Add} method in
\texttt{MessageCenterNotificationManager} (in the C++ code base) to trigger the
display of a notification. Also, we found that \texttt{WebNotificationDelegate}
has a method called \texttt{Click} that is responsible for propagating user
clicks to the notification. Therefore, to simulate a user click we instrument
the \texttt{Add} method mentioned above so that, after the notification is
displayed, it waits for a short delay (e.g., a few seconds) and then calls
\texttt{WebNotificationDelegate::Click}.
If the \texttt{Click} call results in a web page navigation, our browser records
all network requests involved in the process (including all redirections) and
fine-grained details about the rendering of the landing page, which includes
detailed information about any JavaScript code executed in the context of the
landing page. The Service Worker responsible for the notification may also
listen to the click event and may respond by closing the notification shown and
contacting a server to report that a click occurred. We also record all these
events and interactions.

\subsection{Mobile Environment}
We also developed a version of \sysname for Android. Due to some
technical differences between how WPNs are displayed on a mobile OS, compared to
desktop environments, we had to adapt some of the system components to run
specifically on Android. First, at the time when we started building our system
Puppeteer~\cite{Puppeteer} did not appear to support Android Chromium
automation. Only recently there have been online posts in which Puppeteer users
describe how they have been able to ``hack'' their configurations to remotely
control an Android browser. We therefore built our own browser automation
framework that works via the Android Debug Bridge (ADB). The capabilities of our
ADB-based automation framework are limited, but sufficient for enabling data
collection for \sysname. We will explore the use of Puppeteer for \sysname on
Android in future work.

\vspace{3pt}
\noindent
\textbf{Logging Internal Browser Events: }
We compile our instrumented Chromium browser for Android, so that we can collect
intimate details about internal browser events related to WPNs, including
recording information about the related Service Workers and the rendering of the
landing page resulting from clicking on a WPN. Browser logs are sent via the
\texttt{logcat} ADB command to a remote logging machine.
%

\vspace{3pt}
\noindent
\textbf{Interacting with Notifications: }
Unlike on desktop devices, in which WPN messages are displayed by the browser,
on Android device it is Android OS that displays a WPN as a system
notification. Also, unlike on desktop environments, the browser does not need to
be activated for a WPN message to be received, though the browser may be activated
after tapping on a notification (e.g., to navigate to the URL pointed to by the
notification). We therefore had to implement a different mechanism to simulate
user interactions with WPNs on Android.
Specifically, we developed an Android application that leverages Android's {\em
accessibility service}. The accessibility service is aimed to help people with
disabilities in using the device and apps. It is a long-running privileged
system service that helps users process information from the screen and lets
them interact with the content meaningfully in an easy way. Android developers
can leverage the accessibility service API and develop apps that are made aware
of certain events, such as \texttt{TYPE\_VIEW\_FOCUSED} and
\texttt{TYPE\_NOTIFICATION\_STATE\_CHANGED}. Furthermore, the accessibility
service API can also be used to initiate user actions such as click, touch and
swipe.

We install our app with accessibility service permission on an Android physical
device, and use it to interact with every notification event fired by our
instrumented Chromium browser. Whenever a new notification pops up, our
application will automatically swipe down the notification bar and click on the
notification to complete the action, while our instrumented Android browser
produces detailed logs about the consequences of such interactions (e.g.,
loading a new web page).

%% file: data_analysis.tex
\section{Data Analysis Module}
\label{sec:data_analysis}

In this section, we first explain in detail how we mine the collected WPN
messages to identify WPN-based ad campaigns, and later explain how we label
malicious campaigns.

\subsection{WPN-based Ad Campaigns}
\label{sec:adCampaigns}
To identify WPN-based ad campaigns, we mine the dataset of WPN messages
collected by \sysname from a large and diverse set of websites. To distinguish
between generic WPN messages and WPN-based ads, we start by considering the
following intuitions. Roughly speaking, advertisers tend to promote their
products or services on multiple websites, and clicking on a WPN notification
typically leads to a third-party landing page, on a different origin than the
website from which the WPN message was received. On the other hand, non-ad WPN
messages are typically related to alerts (e.g., breaking news, weather alerts,
etc.) that are specific to the notifying website itself, and clicking on the
notification often leads to the same origin to which the notification was
related.

Following the above intuitions, we broadly define a {\em WPN ad campaign} as
{\em a set of WPN messages from multiple sources that deliver similar content
promoting the same (or similar) products or services}. In practice, this
translates into a (potentially large) group of similar WPN messages pushed by
multiple different websites that lead to the same landing page, or different
landing pages that show similar content.

To find such WPN ad campaigns among a large collection of generic WPN
messages, we leverage a document clustering approach. This clustering process
aims to group together WPN messages that are similar to each other in terms of
message content and landing page. As a result, WPN ads that belong to a large
WPN ad campaign will tend to form larger clusters. On the other hand, non-ad WPNs would
tend to be isolated into singleton clusters or clusters that contain
messages related to only one source website and a landing domain that points back to the source website itself.

Because clustering is an unsupervised learning process, it is usually
challenging to tune the hyperparameters to obtain perfect clustering results.
Therefore, to minimize the chances of grouping together ads and non-ads, we tune
our clustering system to be conservative and yield ``tight'' clusters. Namely,
WPN messages assigned to the same cluster will exhibit high similarity. Although
this may result in some WPN ads being left out of the campaign (i.e., the
cluster) they may also belong to, we will see later that we can reconnect them to their respective campaigns using a meta-clustering step (Section~\ref{sec:cluster_groups}).

\subsubsection{WPN Clustering Features and Approach}
\label{sec:wpnclustering}

To cluster similar WPNs, we first need to define the features to be compared and
a similarity function that calculates the closeness of two WPN messages. Thanks
to our instrumented browser, each WPN we collect is accompanied by metadata such
as the notification title, message body, the images and icons found in
the WPN, the URL of the landing page, a screenshot of the landing page reached
after clicking on the WPN (in the case of our desktop browser), 
the events that occur immediately after a notification click, etc.  

Among the above features, to discover WPN ad campaigns we use the following ones
for clustering: message title, message body, and the URL path (i.e., we exclude
the domain name) of the message's landing page URL. We then use these features
to calculate the pairwise distance between WPN messages. Specifically, given two
messages, we compute the distance between their title and body text, and
separately the distance between their landing URL paths. Then, we compute the
total distance between two WPNs as the average of these two distances. We
explain the distance measures in more detail below. The remaining information
contained in the WPN metadata collected by our browser (e.g., domain names,
screenshots, etc.) that are not used as features will instead be used later to
help us validate the clustering results.
    
\textbf{WPN Messages Distance:} The text contained in the title and body of a
WPN message is typically short and includes specific keywords that reflect the
{\em theme} of the message. To measure the similarity between these short pieces
of text (the concatenation of title and body), we require a measure that gives
importance to significant keywords found in the content of WPNs. To this end, we
use the {\em soft cosine similarity}~\cite{softcossim} measure, which considers
the semantic relationship between words. To compute the similarity between
words, we first use Word2Vec~\cite{Word2Vec} on all WPN messages to obtain a term-similarity
matrix.  We then convert each WPN message into a bag-of-words representation,
and input both the term-similarity matrix and the bag-of-word vectors for each
pair of WPN messages into the cosine similarity function (we use {\tt
softcossim()} implemented in {\tt gensim}\cite{gensim}) to obtain a pairwise
similarity matrix for WPN messages. Since the similarity value $s\in[0,1]$, we
calculate the distance as $d = (1-s)$.

\vspace{2pt}
\noindent\textbf{URL Paths Distance:} Given the landing page URL of a WPN (which
we collect along with other metadata, as explained earlier), we extract the URL
path by excluding the domain name and the query string values, while retaining
the relative path to the page and query string parameter names. To calculate the
distance between two URL paths, we use the Jaccard distance between tokens
extracted from the URL path, such as the components of the directory path, the
page name, and the name of the query string parameters. 

\vspace{2pt}
\noindent\textbf{WPN Clustering:} Once the pairwise distances are computed, we
use agglomerative hierarchical clustering over the distance matrix. To determine
where to cut the resulting dendogram, we compute the average silhouette
score~\cite{cluster_validity} for clustering results obtained at different cuts,
and choose the cut with the highest score. We refer to the result as {\em WPN clusters}.

\vspace{2pt}
\noindent\textbf{WPN Ad Campaigns:} As per our definition of WPN ad campaigns,
to determine if a WPN cluster is formed by ads, we take into consideration the
source of the WPNs. Namely, we count the number of effective second-level domain
names associated with the websites that sent the push notifications per each
cluster. This number tells us if the WPNs have been published on multiple
sources. If a cluster contains more than one distinct second-level source
website domain, we label the cluster as an {\em WPN ad campaign}.


\subsection{Identifying Malicious WPN Clusters}
\label{sec:malicious_campaigns}
To determine the maliciousness of a WPN cluster, we leverage two well-known URL
blacklisting services: Google Safe Browsing(GSB)\cite{gsb} and Virus
Total\cite{virus_total} (VT). We submit the full URLs of all the landing pages
reached from all WPN messages in the cluster to these services. Then, we label a
particular WPN message as {\em known malicious} if the landing page URL is
blacklisted as malicious by any of the blacklisting services. Next, we use a
simple label propagation policy to flag as malicious the WPN clusters that
contain at least one known malicious WPN. 

It is worth noting that we submit full URLs to GSB and VT. If a full URL, $u$,
is blacklisted, it does not necessarily mean that all URLs under the same domain
name as $u$ will also be blacklisted (in fact, we found some cases that confirm
this observation for VT). In addition, similar malicious WPN messages often lead
to different domain names, mainly as an attempt to evade blocking by URL
blacklists. At the same time, because WPN messages in the same cluster are very
similar, thanks to our conservative clustering approach, they share very similar
title, body, and structure of the landing page URL path. Intuitively, if one WPN
message is known to lead to a malicious landing page, it is highly likely that
all other WPN messages in the same cluster also lead to similar malicious pages,
as we will also discuss in our measurement results (see
Section~\ref{sec:results}). That is why we rely on the simple ``guilty by
association'' policy mentioned above for label propagation.

\subsection{WPN Meta Clustering}
\label{sec:cluster_groups}
Because URL blacklists have limited coverage, it is possible that some malicious
WPN clusters will not be immediately identified using the labeling approach
discussed above. Furthermore, because our clustering approach is conservative,
it is possible that separate clusters of WPN messages may in fact be related to
each other. To compensate for this, we perform a meta-clustering step that aims
to group together WPN clusters that may belong to the same WPN ``operation''
(e.g., the same advertiser), as explained below.

To this end, we generate a bipartite graph $\mathcal{G} = (W, D, E)$. $W$ is a
set of nodes in which each node represents a WPN cluster obtained as described
in Section~\ref{sec:wpnclustering}. $D$ is the set of all domains pointed to by
the WPN messages we collected (i.e., all domains related to any of the landing
page URLs found in the WPN messages), and $E$ is a set of edges in which each
edge connects a node $w \in W$ to a node $d \in D$. In other words, we connect
each WPN cluster to the domains related to the landing page URLs linked by the
WPN messages in the cluster. Then, we find all isolated components, ${G_1, G_2,
\dots, G_m} \in \mathcal{G}$, and consider each isolated component as a {\em
meta cluster} of WPN messages. Notice that this leads us to groups of WPN
clusters that are related to each other because they collectively share common
landing page domains. Figure~\ref{fig:bi-partite} in
Appendix~\ref{sec:metacluster_graphs} visually shows two examples of such
meta clusters.

\subsection{Labeling WPN Meta Clusters}
\label{sec:double_serving}

First, we consider as {\em suspicious} all domain names associated with full
URLs that have been previously labeled as malicious by GSB or VT. Then, let $G_i
\in \mathcal{G}$ be a meta cluster that includes one or more of such suspicious
domains. We label WPN clusters in $G_i$ (and thus all WPN messages in those
clusters) as suspicious, unless they were previously labeled as malicious
according to the process described in Section~\ref{sec:malicious_campaigns}.

In addition, given a meta cluster $G_i$, if it contains at least one WPN
cluster $w_j \in W$ that we previously identified as a {\em WPN ad campaign}
(see Section~\ref{sec:adCampaigns}), we consider all WPN messages contained in
the WPN clusters within $G_i$ as WPN ads. This is because those WPN messages
point to domain names related to WPN-based advertising, since they are linked to
one or more WPN ad clusters, and thus are highly likely WPN-based ad messages
themselves.

To identify additional suspicious WPN ad campaigns that were not previously
labeled based on the process described above and in
Section~\ref{sec:malicious_campaigns}, we proceed as follows. We notice that ad
networks such as Google Ads and Bing Ads recommend advertisers not to promote
the same product, brand, or similar customer experience by redirecting users to
multiple destination websites. Violating this policy is referred to as {\em
Abuse of Ad Network} by Google Ads\cite{google_ads} and {\em Duplicate Ads} by Bing
Ads\cite{bing_ads}.
Besides the fact that these practices do not conform to advertising policies,
malicious advertisers often prefer hosting similar malicious content on multiple
domains to evade detection and to continue the attack even if one of their
domains is blacklisted. We then identify such practices and label the related
meta-clusters that lead to multiple different landing domains
as {\em suspicious}, to trigger further (manual) analysis. 
We provide a detailed measurements on
the identified suspicious WPN ads in Section~\ref{sec:eval_suspicious_ads}.

\noindent\textbf{Manual Verification:} 
To validate the malicious labels assigned by leveraging URL blacklists, and to
measure the number of suspicious WPN clusters that are in fact malicious, we
manually analyze all malicious and suspicious WPN clusters discovered by our
data analysis module. During manual analysis, we consider multiple factors to
determine if the WPNs are indeed malicious. Once we manually confirm that a WPN
cluster is malicious, we add it to a {\em manual blacklist}, which we then use
to inform the measurement results presented in Section~\ref{sec:results}. 

Some of the factors considered during manual blacklisting are as follows. We
recognize a WPN message as malicious if it meets a combination of the following
conditions: (1) leads to a landing page visually
similar to a known malicious page (as determined by GSB and VT); (2) contains
the same WPN message as a known malicious WPN message, but leads to a different
product/site (i.e., a different landing page); (3) includes message content that
is highly likely malicious or leads to a page with likely malicious content such
as rewards that clearly seem too good to be true, as is typical of survey
scams~\cite{survey_scam}, false financial alerts, etc.; or (4) leads to a
landing page that shares several domain-related properties with known malicious
sites, such as IP address, registrant, very similar domain name, etc. We further
discuss our manual analysis process by presenting some examples in
Section~\ref{sec:eval_malicious_campaigns}.

%% file: evaluation.tex
\section{Measuring WPN Ads in the Wild}
\label{sec:results}
In this section, we report measurements on the usage of WPNs as an ad delivery
platform, and provide insights into the malicious use of WPN ads. 

\input{collection_setup.tex}

\input{tables/results.tex}

\subsection{WPN Messages Dataset}
\label{sec:notifications}
We start with the 5,849 initial URLs that we collected as explained in
Section~\ref{sec:crawler_seeding}, over 5,697 distinct second-level domain names.
By clicking on WPN messages issued by these initial URLs, we collect an
additional 10,898 URLs across 2,269 distinct second-level domains. When visited,
many of these additional URLs presented our browser with a notification request,
which our crawler automatically granted. This brought us to a total of 7,951
URLs that registered a Service Worker with Push permission and were therefore
able to push notifications to our instrumented browser instances over time. 

During the course of about two months (September and October 2019), we were able
to collect a total of 21,541 push notification messages, including 12,441
notifications for the desktop environment and 9,100 for the mobile environment.
\sysname interacted with each of these WPN messages. However, not all automated
clicks on notification boxes led to a separate landing page. In addition, some
landing pages appeared to cause a crash in the browser's tab (but not the
overall browser) in which they rendered, preventing us from collecting detailed
information on those pages (this was likely due to the fact that our
instrumented Chromium browser is not based on the most recent stable code base).
We filtered out these notifications, because we could not collect sufficient
details about the related landing pages, leaving us with 12,262 WPN messages
(9,570 on desktop and 2,692 on mobile) that when clicked on lead to a valid
landing page. We then used this final set of WPN messages in the clustering
process described in Section~\ref{sec:data_analysis}.

\subsection{Data Analysis Results}

\noindent\textbf{Summary of findings}: Table~\ref{tab:cluster_results}
summarizes the overall results of our analysis process. From the 12,262 WPN
messages mentioned above, \sysname identified 572 WPN ad campaigns and a total
of 5,143 WPN ads related to these campaigns. Moreover, \sysname identified 51\%
of all WPN ads as malicious. Specifically, \sysname found 318 (out of 572)
campaigns to be malicious; in aggregate, these malicious campaigns included
2,615 WPN ads.

This is quite a staggering result, in that it appears that ad networks that
provide WPN ad services are heavily abused to distribute malicious content.
Later, in Section~\ref{sec:ad_blockers}, we also show that ad blockers are
ineffective at blocking such ads, which is an additional cause of concern. In
the following sections, we discuss the clustering and labeling results in more
details.

\subsubsection{WPN Clusters and Ad Campaigns}
\label{sec:cluster_campaigns}
As discussed in section \ref{sec:wpnclustering}, we cluster the collected WPN
messages based on their message content and landing page information. After clustering 12,262 WPN messages that led to a valid landing page, we obtained 8,780 WPN clusters, of which 7,731 were singleton clusters containing only one element (i.e., only one WPN message). Of the remaining non-singleton
clusters, 572 were labeled as WPN ad campaigns, according to the process
described in Section~\ref{sec:data_analysis}. In aggregate, these WPN ad
campaigns pushed 3,213 WPN ad messages to our browsers, during a period of about
two months. We now provide a few examples of WPN clusters, and discuss what type
of WPNs tend to fall within singleton clusters. 

\begin{figure*}[!ht]
\setlength\belowcaptionskip{-0.7\baselineskip}
\centering
\includegraphics[scale=0.48]{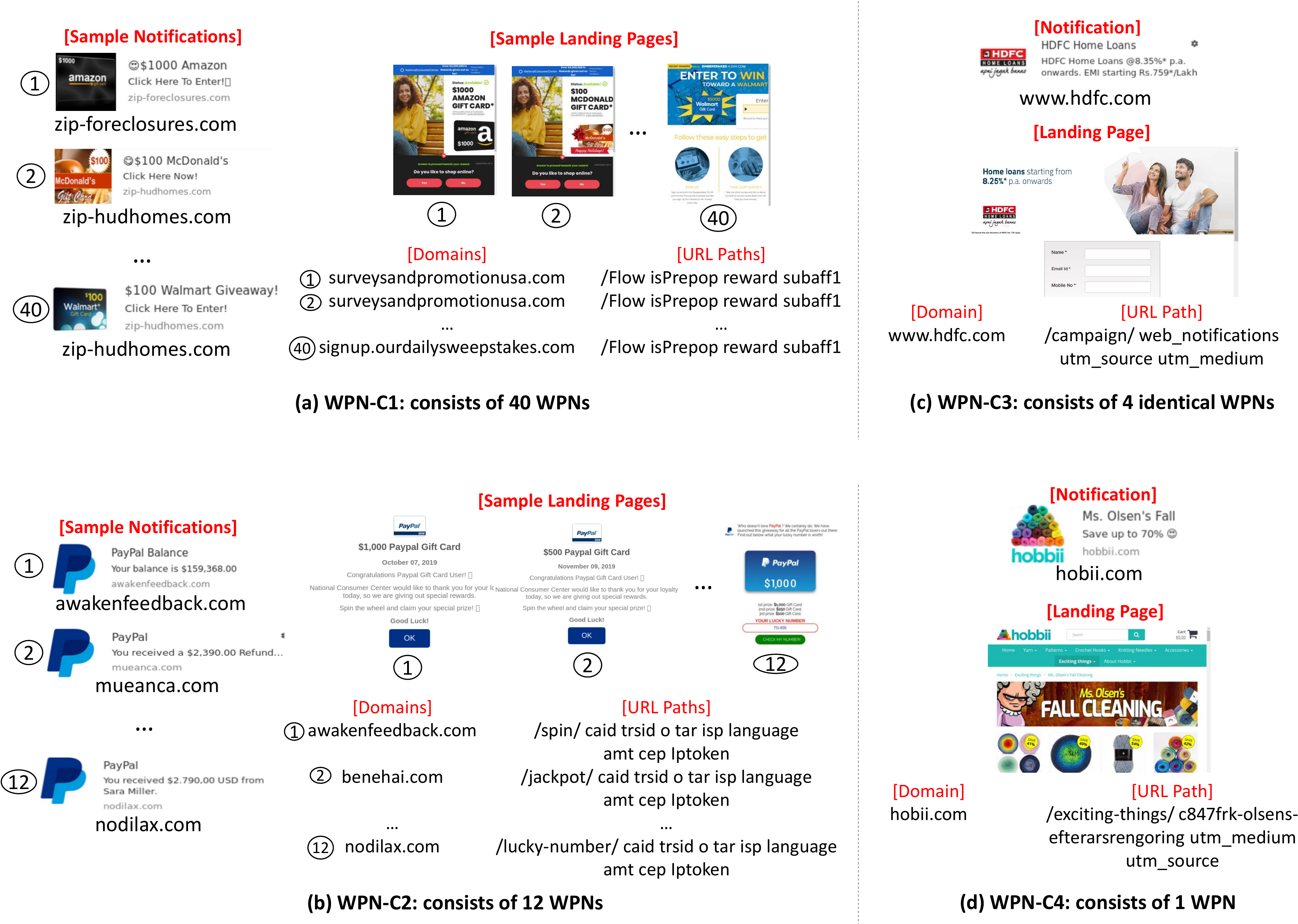}
\caption{Examples of WPN clusters}
\label{fig:wpn_clusters}
\end{figure*}

Figure~\ref{fig:wpn_clusters} provides some examples of WPN clusters. 
Cluster {\tt WPN-C1} consists of 40 WPNs that advertise free giveaways
and prizes, which we found led to pages requesting users to complete surveys
or provide detailed personal information to claim the offer (this is a typical
survey scam campaign, as discussed in
Section~\ref{sec:eval_malicious_campaigns}). {\tt WPN-C2} consists of 12 WPN
messages that alert users about a large increase in their Paypal account
balance. These WPN messages led to pages that offer users to claim their money
after providing personal information (another form of scam). In both {\tt
WPN-C1} and {\tt WPN-C2}, the respective WPNs were pushed from multiple
sources (i.e., multiple second-level domain names), as also shown in
Figure~\ref{fig:wpn_clusters}.

{\tt WPN-C3} included 4 identical WPN messages pushed by a single source
website, a bank, alerting users on their loan offers. These messages appear to
be legitimate, and led back to the site that pushed them. {\tt WPN-C4} is an example of WPN message isolated into a singleton cluster. This message was isolated during the clustering process because
they carry different content and/or led to different URL paths,
compared to other WPN messages.

According to the definition provided in Section~\ref{sec:data_analysis}, we
label {\tt WPN-C1} and {\tt WPN-C2} as WPN ad campaigns, because the WPNs in each of the clusters deliver very similar (or the same) message
promoting very similar products from multiple sources. However, {\tt
WPN-C3} and {\tt WPN-C4} do not meet the definition and are thus
not labeled as WPN ad campaign.

\subsubsection{Malicious WPN Ad Campaigns}
\label{sec:eval_malicious_campaigns}
As described in Section \ref{sec:malicious_campaigns}, we submit landing page
URLs related to all WPN messages to GSB~\cite{gsb} and VT~\cite{virus_total}. On
our initial scan, less than 1\% of the URLs were detected as malicious by GSB or
VT, in aggregate. For instance, initially VT flagged 108 landing page URLs as
malicious, of which 88 were related to WPN messages labeled by our system as
belonging to ad campaigns. Notice that for VT we consider a URL as malicious if
at least one of the URL detection engine reports it as malicious, and later
manually review all results to filter out possible false positives. After one
month, we submitted the same set of URLs once again, and we found that 1,388
(11.31\%) of them were then detected by VT, though GSB still only flagged 1\% of
them. 

\sysname relies on label propagation to label WPN messages and clusters as
malicious, based on results from VT and GSB, as explained in
Section~\ref{sec:malicious_campaigns}. To limit the chances of amplifying
possible false positives from VT and GSB, we manually verified all 1,388 URLs to
check whether they actually led to malicious content. We were able to confirm
that 96.8\% of them indeed appeared to be malicious. Of the remaining 44 URLs
that we could not confirm as malicious, 13 were found to belong to popular
benign domains such as \url{bing.com}, \url{kbb.com}, \url{tophatter.com}, etc.;
24 URLs were related to unpopular blog/news sites; 3 led to adult websites; and
4 led to websites hosting non-English content that we could not
verify. Given that these sites may be benign, since we do not have all the
information VT and GSB had to label them as malicious we take a conservative
stance and remove the malicious label from them.  Accordingly, we label 1,344
WPNs as {\em known malicious}. Among them, 758 WPNs were part
of 572 WPN clusters that we previously classified as ad campaigns (see
Section~\ref{sec:cluster_campaigns}). The remaining 586 WPN messages that led to
malicious landing pages were not immediately found to belong to WPN ad clusters,
as they formed separate small clusters. We will determine whether they are
related to WPN campaigns later, in Section~\ref{sec:eval_suspicious_ads}) after
the meta-clustering step (explained in Section~\ref{sec:cluster_groups}).

\input{tables/detection.tex}

By using a ``guilty by association'' label propagation policy, as explained in
Section~\ref{sec:malicious_campaigns}, we label WPN ad campaigns as malicious if
they include at least one {\em known malicious} WPN (remember that this
policy is justified by the close similarity in content and landing page URL path
between messages in the same cluster). This yielded 152 (out of 572) {\em
malicious} WPN ad campaigns, which overall included 376 WPN 
(or more precisely their landing pages) that GSB or VT missed to detect as
malicious. After manually inspecting these 376 WPN ads, we were able to confirm
that 367 of them are indeed malicious ads that lead to survey scams, phishing
pages, scareware, fake alerts, social media scams, etc. We were not able to
confirm the maliciousness of the remaining 9 ads (i.e., 2.4\%) that led to
different pages that welcome/thank the user for subscribing to the notification
all hosted on the same IP address. 
The take away from the above discussion is that, using our WPN clustering approach, we were able to increase the
number of confirmed malicious WPN ads from 758 to 1,125 (i.e., 758 plus 367),
which represents an increase of about 50\%. The above results are summarized in
Table~\ref{tab:malicious_results}, first row.

Referring back to the examples provided in Figure~\ref{fig:wpn_clusters},
in cluster {\tt WPN-C1}, 35 out of the 40 WPNs
were labeled as {\em known malicious} WPNs, according to VT. As mentioned
earlier, \sysname labeled this entire cluster as {\em malicious}. After manually
inspecting all 40 messages, we confirmed that the remaining 5 messages in the
cluster were indeed related to the 35 malicious sweepstakes/survey scam ads.

\subsubsection{Finding Suspicious Ads}
\label{sec:eval_suspicious_ads}

So far, we have leveraged the labels provided by VT and GSB to identify
malicious WPN ads, and label WPN ad campaigns. Unfortunately, both URL
blacklists suffer from significant false negatives, when it comes to detecting
malicious landing pages reached from WPN ads. As an example, consider cluster
{\tt WPN-C2}, which \sysname identifies as an ad campaign. This cluster contains
12 WPNs; none of which were labeled as {\em known malicious} according to
VT. However, \sysname flags this cluster as suspicious since it contains 'Duplicate Ads' and via manual inspection we found that the WPN messages in this cluster display fake PayPal alerts that lead users to survey scam pages;
therefore, we manually label the entire {\tt WPN-C2} cluster as malicious. This
example demonstrates the gaps left by current URL blacklisting services, and how
ineffective they could be if they were used to detect and block malicious ad
notifications. Below we discuss how we use the meta-clustering
approach explained in Section~\ref{sec:double_serving} to automatically identify
and label more of such cases.

As described in Section \ref{sec:cluster_groups}, we apply a meta-clustering
method to group WPN clusters that may relate to each other, as they share common
landing page domains. To this end, we create a bipartite graph $\mathcal{G} =
(W, D, E)$, here $W$ is the set of all 8,780 WPN clusters we previously
obtained, and $D$ is the set of all 2,177 distinct landing page domains pointed
to by WPN ads that we were able to record. By identifying and separating the
connected components in this bipartite graph, we identify 2,046 WPN meta
clusters. Of these, 224 contain a mix of WPN clusters that we previously labeled
as {\em ad campaign} and other non-campaign WPN clusters. We then label all WPN
messages contained in these 224 ad-related meta-cluster as {\em WPN ads}, thus
increasing the number of WPN ads identified so far from 3,213 to 5,143. More
specifically, let $G_i = (W_i, D_i, E_i)$ be a connected component subgraph of
$\mathcal{G}$. If at least one WPN cluster $w_k \in W_i$ was previously labeled
as {\em WPN ad campaign} (see Section~\ref{sec:adCampaigns}), we label the
entire meta-cluster $G_i$ as ad-related, and thus consider all WPN
messages contained in the union of cluster $\bigcup_k w_k$ as {\em WPN ads}. Figure~\ref{fig:bi-partite} provides two examples of meta clusters, which are
discussed in Appendix~\ref{sec:metacluster_graphs}, due to space constraints. 

Next, we consider all yet to be labeled WPN messages in a WPN meta cluster as
suspicious if the meta cluster contains at least one {\em malicious} WPN cluster
or if it contains {\em duplicate ad domains}, as defined in
Section~\ref{sec:double_serving}. Out of the 572 WPN ad campaigns identified
earlier, we found that 255 of them contained {\em duplicate ad domains}.
Accordingly, we were able to label a total of 287 out of 2,046 WPN meta clusters as
{\em suspicious}. Further, we identified 166 (out of 572) additional WPN ad
campaigns, which were not previously labeled malicious in the previous step, as {\em suspicious}. Overall, this translates into 1,479 suspicious WPN ads, as shown in
Table~\ref{tab:malicious_results}. Following our manual verification process, we
confirmed 1,280 (86.5\%) of these ads as malicious. The remaining 199 WPN ads
were flagged by \sysname because they were related to {\em duplicate ad
domains}. Of these, 166 were alerts related to job postings and led to similar
pages on multiple domains listing the same job; 23 led to multiple sites that
hosted content related to the horoscope; 4 led to adult websites; and 6 were
subscription welcome/thank you notifications.  Notice that while these 199 WPN
messages may be benign, \sysname helped us identify and characterize a large
number of additional WPN ads that are in fact malicious and were not identified
by URL blacklists such as VT or GSB. However, our current system is not designed
to be an automatic malicious WPN ad detection system. In our future work, we
plan to leverage the lessons learned from the measurement results obtained in
this paper to investigate how malicious WPN messages can be accurately detected
and blocked in real time.

\subsection{Ad Networks and Ad Blocking Effects}
\label{sec:ad_blockers}
\begin{figure}[h]
\vspace{-4mm}
\setlength\belowcaptionskip{-0.7\baselineskip}
\setlength\abovecaptionskip{-0.2\baselineskip}
\centering
\includegraphics[width=\columnwidth]{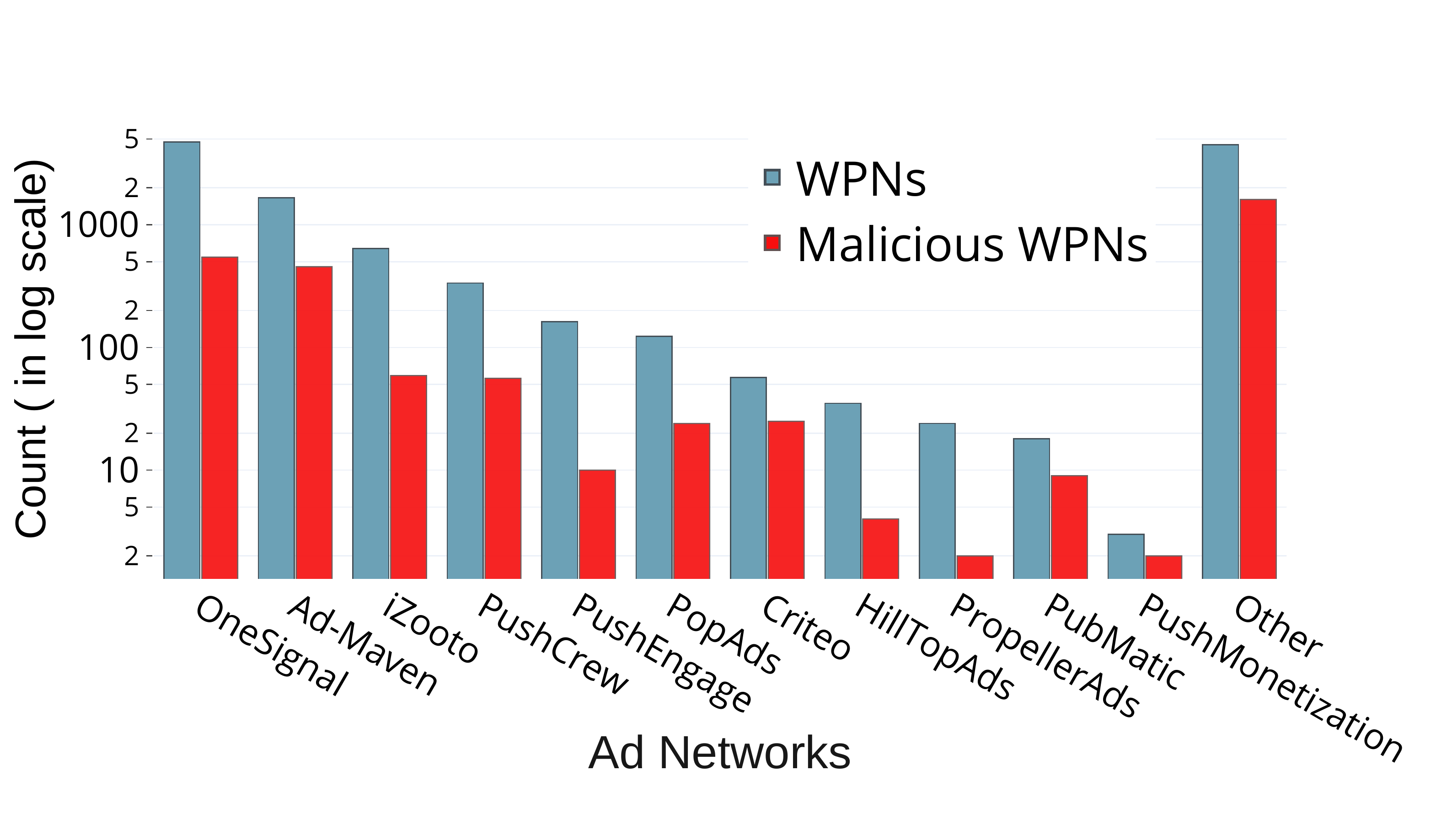}
\caption{Distribution of WPNs w.r.t. Ad Networks}
\label{fig:ad_networks}
\end{figure}

Figure~\ref{fig:ad_networks} shows the distribution of WPN ads, including
malicious ones, per ad network. As it can be seen, many of the ad networks we
considered in our measurements are abused to distribute malicious WPN ads. 

We also investigated whether ad blocker extensions used by desktop browsers may be effective against WPN ads.
Unfortunately, because browser extensions have limited visibility into Service
Worker activities~\cite{master_web_puppets}, they are currently unable to
mitigate the rise of WPN ads. For instance, we checked the URLs of Service
Worker scripts against the Easylist filter rules\cite{easylist} used by most
popular ad blockers. Furthermore, we installed two highly popular ad blocker
extensions in our Chromium browser and checked its blocking capability.  As
shown in Table~\ref{tab:adblockResults}, both ad blocking mechanisms failed to
block the registration of Service Worker scripts related to ad networks that
support WPN ads, even though Easylist was able to filter a small number
(less than 2\%) of network requests subsequently issued by the installed Service
Worker scripts. This shows that existing methods are not sufficient to mitigate
WPN-based ads, including malicious ones.

\input{tables/adblock.tex}

%% file: collection_setup.tex
\subsection{Data Collection Setup}
\label{sec:data_collection_setup}

We first describe \sysname's setup for harvesting in-the-wild WPN messages for
both desktop and mobile environments. Because our internal browser
instrumentations are implemented by extending the browser code provided
by~\cite{Secma}, our data collection process leverages Chroumium's code base
version 64.0.3282.204, which we built for both Linux and Android environments.
As mentioned earlier, we plan to release the source code of our system, along
with the WPN messages dataset collected for this study.

\subsubsection{Seeding the WPN Crawler}
\label{sec:crawler_seeding}
Our main goal is to setup the WPN data collection system to maximize our chances
of collecting WPN-based ads, so that we can measure their properties and
discover abuse. To this end, we first identify a {\em seed} set of popular
advertisement networks that support push notification advertisements.
Specifically, by manually performing online searches and reading articles and
forums dedicated to advertising online, we manually discovered 15 popular ad
networks that provide push advertisement services. We registered an account with
9 of these ad networks to obtain the JavaScript code that needs to be embedded
in an ad-publishing websites to include the ad networks' Service Workers that
will control their push notifications requests. We then used the
\url{publicwww.com} code search engine to find websites (i.e., URLs) that likely
embed the identified ad network's code. The code search engine,
\url{publicwww.com} itself provides a list of {\em ad networks to search
keywords} mappings~\cite{publicwww_exad}, which we leveraged to obtain URLs of
websites that likely embedded ads from the remaining 6 seed ad networks of
interest, without the need to register an account with those networks as well.
The list of 15 seed ad networks we discovered this way is shown in
Table~\ref{tab_ad_url_counts}. 

In addition to the search keywords related to the 15 ad networks mentioned
above, we further used generic keywords that potentially indicate that a web
page employs the Push Notifications: {\tt NotificationrequestPermission},
{\tt pushmanagersubscribe}, {\tt addEventListener('Push'}, {\tt adsblockkpushcom}.
Thus, overall we obtained a list of 19 code search keywords to be used for
searches on \url{publicwww.com} that are likely to lead to URLs that make use of
push notifications. Table~\ref{tab_ad_url_counts} provides a count of the HTTPS
URLs found via \url{publicwww.com} for each ad network and generic WPN-related
keyword.

As a result of the above search, we were able to gather a total of 87,622
HTTPS URLs of potential WPN ad publishing web pages. These URLs were hosted on
82,566 distinct second-level domain names.
We use this set of URLs as seed for \sysname's WPN message crawlers.
%
%
Notice that there is no guarantee that the sites we crawl will in fact lead us
to registering for and receiving WPN-based ads. Rather, only a subset of these
web pages will actually request notification permissions. Therefore, we visited
each URL and retained only those that actually make a request for a notification
permission. Then, we used \sysname to automatically grant notification
permission requests on those URLs. As shown in the last column of
Table~\ref{tab_ad_url_counts}, overall we identified 5,849 URLs hosted on 5,697
distinct second-level domains that issued a notification permission request.

\subsubsection{Collecting WPNs in a Desktop Environment}
\label{sec:desktop_setup}

To automatically harvest WPN advertisements at a large scale, we leverage Docker
containers~\cite{docker} to launch several parallel instances of our
instrumented browser-based crawlers. We observe that a few ad networks use
cookies or other information to track the device or web browser across browsing
sessions. To mitigate this and increase the chances of being presented with
notification permission requests from as many sources as possible, we create a
separate docker container for each URL we visit. During our experiments, we used
four different Ubuntu 16.04 Linux machines having between 8 to 32 CPU cores and
64 to 128 GB of memory each, running a total of 20 to 50 docker sessions in
parallel at a time. For each seed URL that issues a notification permission
request, we start the monitoring phase described below.

Every time we visit a URL, we wait 5 minutes to make sure the website has
sufficient time to present our browser with a notification request. Once a
permission request is received and automatically granted by a browser instance,
a Service Worker is registered. If a Service Worker was registered, we keep the
related Docker container alive for an additional 15 minutes, to allow the
browser to receive the first (or more than one) WPN message from the visited
URL. To select this 15 minutes threshold, we first performed pilot experiments
with much longer waiting times (up to 96 hours) for a large subset of URLs
(1,425 URLs, to be precise), and observed that 98\% of them sent their first
notification within 15 minutes of when the permission was first granted. 

Given a container, and therefore a browser instance that granted notification
permissions to a specific URL, after the first 15 minutes of its life we suspend
the container to free up resources for instantiating a new container that will
visit a new URL. However, we periodically resume suspended containers to see if
they will receive additional notifications, which are queued in the FCM and sent
to the browser as it comes back online.

\input{tables/domains.tex}

\subsubsection{Collecting WPNs in Mobile Environment} 
\label{sec:mobile_setup}

During our study, we found that WPN messages sent to mobile devices tended to be
somewhat different that the ones collected by desktop browsers, in that they
were more tailored to mobile users. In particular, malicious mobile WPN messages
included fake {\em missed call} notifications, fake {\em amber alerts},
``spoofed'' {\em Gmail} or {\em WhatsApp} notifications, fake {\em FedEx}
notifications, etc. In addition, we found that these malicious messages were
much more likely to appear on real Android devices, rather than emulated
environments (likely due to some form of emulator detection). Therefore, to
automatically collect mobile WPN messages we instrumented a real mobile device.
Specifically, we used a Google Nexus 5 device with 2 GB of RAM and a
1080$\times$1920 pixels display. The Android version we used was  {\tt
aosp\_shamu-userdebug 7.1.1 N6F26Y}.

As we attempted to scale our \sysname's mobile WPN crawlers on a real device, we
identified two challenges. First, Docker or other container techniques do not
support Android, and therefore we cannot easily visit multiple URLs in parallel
with isolated browsing sessions. Second, we considered to use app cloning
techniques~\cite{ParallelSpace} to open multiple browser instances separately in
isolated execution environments. However, the limited computing power of our
mobile device restricted us to scale up the experiments and visit a large number
of URLs simultaneously. Therefore, we decided to open multiple URLs in one
chromium app but in separate tabs. While we might miss some advertisements that
use device or web browser tracking techniques, we found that the number of WPNs
we were able to collect from each registered Service Worker is 3.5 times more
than the desktop environment on average. In addition, we disable third party
cookies, data saver, and other options that might interfere with our data
collection.

%% file: tables/domains.tex
\begin{table}[htbp]
    \setlength\belowcaptionskip{-0.4\baselineskip}
    \setlength\abovecaptionskip{0.2\baselineskip}
    \caption{URLs and Notification Permission Request counts}
    \centering
    \scriptsize
    \begin{tabular}{l|c|c}
    \toprule
    \hline
    {\textbf{Ad Network}} & \centering\textbf{URLs}  & \textbf{\makecell{NPRs}}\\
    \hline
     Ad-Maven & 49,769  & 1,168
     \\
     \hline
     PushCrew & 15,177  & 427
     \\
     \hline
     OneSignal & 11,317  & 2,933
     \\
     \hline
     PopAds & 1,582  & 73
     \\ 
     \hline
     PushEngage & 796  & 215
     \\
     \hline
     iZooto & 676  & 278
     \\
     \hline
     PubMatic & 647  & 7
     \\
     \hline
     PropellerAds & 335  & 9
     \\
     \hline
     Criteo & 154  & 5
     \\
      \hline
     AdsTerra & 115  & 2
     \\
      \hline
     AirPush & 52  & 0
     \\
      \hline
     HillTopAds & 21  & 3
     \\   \hline
     RichPush & 12  & 0
     \\   \hline
     AdCash & 10  & 0
     \\   \hline
     PushMonetization & 9  & 5
     \\  \hline
    
     \bottomrule
     {\textbf{Generic Keywords}} & \centering\textbf{URLs}  &
     \textbf{\makecell{NPRs}}\\
     \hline
     NotificationrequestPermission & 3,965 & 538
     \\   \hline
     pushmanagersubscribe & 2,667  & 158
     \\  \hline
     addEventListener('Push' & 263 & 9
     \\  \hline
     adsblockkpushcom & 55 & 19
     \\  \hline
     \bottomrule
     {\bf Total} & 87,622  & 5,849
     \\   \hline
    \end{tabular}
    
    \label{tab_ad_url_counts}
    \vspace{-5mm}
\end{table}

%% file: tables/results.tex
\begin{table}[htbp]
\setlength\belowcaptionskip{-0.5\baselineskip}
\caption{Measurement Results of Data analysis Module}
\label{tab:cluster_results}
\resizebox{\linewidth}{!}{%
\begin{tabular}{c|c|c|c|c|c}
\toprule
\hline
\multicolumn{1}{c|}{\textbf{}} & 
\multicolumn{1}{c|}{\textbf{\begin{tabular}[c]{@{}c@{}}WPNs \\ with\\ Landing \\ Pages\end{tabular}}} & 
\multicolumn{1}{c|}{\textbf{\begin{tabular}[c]{@{}c@{}}WPN \\ Ad\\ Campaigns \\ \end{tabular}}} &  
\multicolumn{1}{c|}{\textbf{\begin{tabular}[c]{@{}c@{}}WPN \\ Ads\end{tabular}}} & 
\multicolumn{1}{c|}{\textbf{\begin{tabular}[c]{@{}c@{}}Malicious \\ WPN Ad \\ Campaigns\end{tabular}}}  & 
\multicolumn{1}{c}{\textbf{\begin{tabular}[c]{@{}c@{}}Malicious \\ WPN \\ Ads\end{tabular}}} 
 \\ \hline
\textbf{Desktop} & 9,570 & \multirow{3}{*}{{572}} &\multirow{3}{*}{{5143}}  & \multirow{3}{*}{{318}} & \multirow{3}{*}{{2615}} 
\\   \cline{1-2}
\textbf{Mobile} & 2,692 &    &  &  
\\ \cline{1-2}
\textbf{Total} & {12,262}  &     &  &  
\\  
\hline
\bottomrule
\end{tabular}}
\vspace{-5mm}
\end{table}

%% file: tables/detection.tex
\begin{table}[ht]
    \setlength\belowcaptionskip{-0.5\baselineskip}
    \setlength\abovecaptionskip{0.4\baselineskip}
    \caption{Measurement Results at Stages of Clustering }
    \label{tab:malicious_results}
    \resizebox{\columnwidth}{!}{%
    \begin{tabular}{l|c|c|c|c|c}
    \toprule
    \hline
    {}
    & \textbf{\begin{tabular}[c]{@{}l@{}}\# clusters\end{tabular}}
    & \textbf{\begin{tabular}[c]{@{}c@{}}\# ad-related \\ clusters\end{tabular}} 
    & \textbf{\begin{tabular}[c]{@{}c@{}}\# WPN \\ ads\end{tabular}} 
    & \textbf{\begin{tabular}[c]{@{}c@{}}\# known\\ malicious ads\end{tabular}} 
    & \textbf{\begin{tabular}[c]{@{}c@{}}\# additional\\ malicious ads\end{tabular}}  
    \\ \hline
    
    \centering\textbf{\begin{tabular}[c]{@{}l@{}}After WPN \\ Clustering \end{tabular}}
    & \begin{tabular}[c]{@{}c@{}}8780 \end{tabular}
    & \begin{tabular}[c]{@{}c@{}}572\end{tabular} 
    & \begin{tabular}[c]{@{}c@{}}3213\end{tabular}  
    & \begin{tabular}[c]{@{}c@{}}758\end{tabular}  
    & \begin{tabular}[c]{@{}c@{}}367\end{tabular}             
    \\ \hline
    
    \centering\textbf{\begin{tabular}[c]{@{}l@{}}After Meta\\ Clustering\end{tabular}}  
    & \begin{tabular}[c]{@{}c@{}}2046  \end{tabular}                                      
    & \begin{tabular}[c]{@{}c@{}}224\end{tabular}                               
    & \begin{tabular}[c]{@{}c@{}}1930\end{tabular}  
    & \begin{tabular}[c]{@{}c@{}}210\end{tabular} 
    & \begin{tabular}[c]{@{}c@{}}1280\end{tabular}            
    \\ \hline
    
    \multicolumn{2}{c|}{}  
    & {{\bf Total:}}                           
    & \begin{tabular}[c]{@{}c@{}}5143\end{tabular}  
    & \begin{tabular}[c]{@{}c@{}}968\end{tabular} 
    & \begin{tabular}[c]{@{}c@{}}1647\end{tabular}            
    \\ \cline{3-6}
    
    \end{tabular}}
\end{table}

%% file: tables/adblock.tex
\begin{table}[htbp]
\setlength\belowcaptionskip{-0.5\baselineskip}
\setlength\abovecaptionskip{0.3\baselineskip}
\caption{Results on Existing Ad Blockers }
\centering
\label{tab:adblockResults}
\resizebox{\linewidth}{!}{%
\begin{tabular}{p{3.5cm}|p{2.5cm}|p{2.5cm}}
\toprule
\hline
\multirow{2}*{} & \multicolumn{2}{c}{\textbf{No. of Blocked URLs}} \\
\cline{2-3}
 & \textbf{Service Worker Scripts} & \textbf{Service Worker Requests} \\
\hline

 Easylist Blacklist & 0 out of 1187 & 132 out of 8031 
\\
\hline
 AdBlockPlus Ad blocker & 0 out of 884 & 0 out of 7276
\\
\hline
 AdGuard Ad Blocker & 0 out of 895 & 0 out of 7520
\\
\hline
\bottomrule
\end{tabular}
}
\vspace{-5mm}
\end{table}

%% file: related.tex
\section{Related Work}
\vspace{3pt}
In this section, we discuss prior studies related to our work.

\noindent\textbf{Service Workers and WPNs:} 
Papadopoulos et al.~\cite{master_web_puppets} propose multiple attack techniques
to register malicious Service Workers that can control a victim's web browser
(e.g.,  steal computation power for performing nefarious activities). There
exist other studies \cite{rushanan2016malloryworker, pan2016assessing} that
demonstrate similar attacks on Web Workers. Lee et al. \cite{pride} study the
risks posed by PWAs. For instance, they collect Service Worker scripts from
top-ranked website homepages and analyze the risks for phishing due to the use
of brand logos in push notifications. They also found a number of push-based
phishing attacks that exploit WhatsApp and YouTube icons. Furthermore, they also
identify security flaws in the design of browsers and third-party push
notification libraries. Other works related to Progressive Web
Apps\cite{pwa_2017,pwa_2018,pwa_energy} focus on the performance of PWAs in
multiple environment and their energy usage. 

Our study is different, because we focus on measuring the use of WPN messages as
an ad-delivery platform in general, and on measusing abuse in WPN ads. To that
end, we design and implement an automated system to subscribe to, collect,
interact with, and analyze in-the-wild WPN messages on both desktop and mobile
platforms.

\noindent
\textbf{Analyzing Online Ads:} 
A line of studies~\cite{bashir2016tracing, barford2014adscape, RafiqueGJHN16} is
dedicated to investigate the online advertisement ecosystem. They trace the
information passed between ad exchanges and analyze the revenue collected using
ads. Another body of studies \cite{MadTracer, masri2017automated, web_malvert_1,
web_malvert_2, web_malvert_3, madlife, madfraud, decaf,
are_ads_safe,starov2018betrayed} focus on identifying {\em traditional}
malicious web advertisements and ad campaigns in both desktop and  mobile
environments. Similarly, a recent study~\cite{Secma} reports observations of
social engineering attack campaigns delivered through traditional web-based
advertisements. Our work is different, because we specifically focus on ads and
malicious ad campaigns delivered through WPN messages.

\noindent
\textbf{Online Scams and Ad Blockers:} 
A number of approaches~\cite{ad_block1,ad_block2, ad_block3} have been proposed
to study the effectiveness of existing ad block techniques as well as the
various counter-measures used by ad providers to circumvent ad blockers. Other
studies~\cite{ipad_scam,survey_scam,dial_one_scam} discuss various online scam
techniques and the prevalence of Internet fraud. Although \sysname does not
focus on blocking malicious ads or online scams, we demonstrate that existing ad
blocking techniques are not effective at defending against malicious WPN ads.

%% file: discussion.tex
\section{Discussion and Limitations}
\label{sec:limitations}

\noindent
{\bf Blocking Malicious WPN Ads:} In Section~\ref{sec:ad_blockers}, we showed
that existing popular ad blockers and filtering rules are not effective at
mitigating WPN ads, due to their limited visibility on Service Worker activities.
An exception is perhaps represented by a browser extension called
AdBlaster~\cite{adblaster}, which particularly claims to block WPN ads. However,
AdBlaster simply disables {\it all} push notifications, including benign non-ad
ones, which can obviously disrupt legitimate uses of WPN messages. Although our
\sysname system is not currently designed to detect malicious WPN ads in a fully
automated way, we believe that the results from our study provide useful
insights into the malicious use of WPN ads that may encourage and help the
security research community to build better defenses against the abuse of web
push notifications.

\noindent
{\bf Handling Permission Request Redirections:} In this study, we observed some
websites that do not directly ask for permission to send notifications to the
browser. Rather, they first create a dynamic JavaScript-based prompt that mimics
a browser permission request. If the user clicks on the confirmation button, the
site then redirects the user to a new page where the true permission to send
notifications is asked through the browser's permission request dialog. It is
possible that this two-step permission granting process, though somewhat more
cumbersome, may be used by some websites to avoid a website being permanently
blocked from requesting notificiation permissions in the future, or to bypass
some other policy restrictions.

Our current version of \sysname is not able to identify ``fake''
JavaScript-driven permission requests invoked by dynamic dialog boxes, and we
therefore potentially miss to collect WPN messages from websites that use the
two-step permission request mechanism described above. In our future work, we
plan to extend \sysname by adding a module that can automatically identify and
interact with such dynamic dialog boxes, for instance by leveraging machine
learning methods. This will allow us to collect additional WPN messages that can
be analyzed by \sysname's data analysis module without requiring any other
significant changes.

%% file: conclusion.tex
\section{Conclusion}

In this paper, we have studied how web push notifications (WPNs) are being used
to deliver ads, and measured how many of these ads are used for malicious
purposes. To enabled this study, we developed a system called \sysname, which
allowed us to automatically collect and analyze 21,541 WPN messages across
thousands of different websites. Among these, our system identified 572 WPN ad
campaigns, for a total of 5,143 WPN-based ads, of which 51\% are malicious. We
also found that traditional ad-blockers and malicious URL filters are remarkably
ineffective against WPN-based malicious ads, leaving a significant abuse vector
unchecked.

%% file: appendix.tex
\begin{appendices}

\section{Meta-Clustering Graphs}
\label{sec:metacluster_graphs}

\begin{figure*}[h!]
    \centering
    \begin{subfigure}{0.4\textwidth}
        \centering
        \includegraphics[scale=0.45]{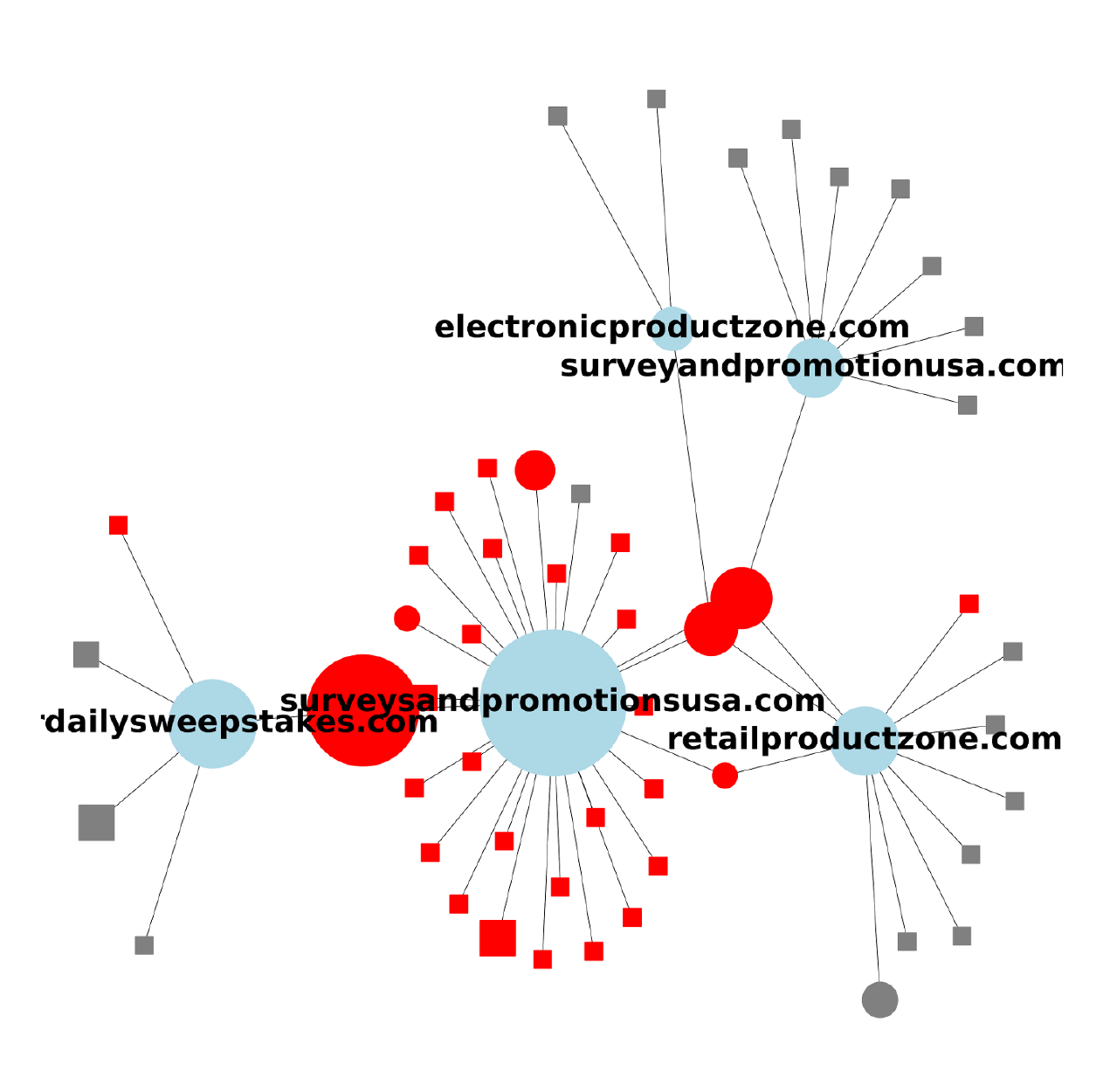}
        \setlength\abovecaptionskip{0.2\baselineskip}
        \caption{Meta Cluster of WPN-C1 WPNAC}
        \label{fig:suspicious_ads_1}
    \end{subfigure}%
    \begin{subfigure}{0.4\textwidth}
        \centering
        \includegraphics[scale=0.45]{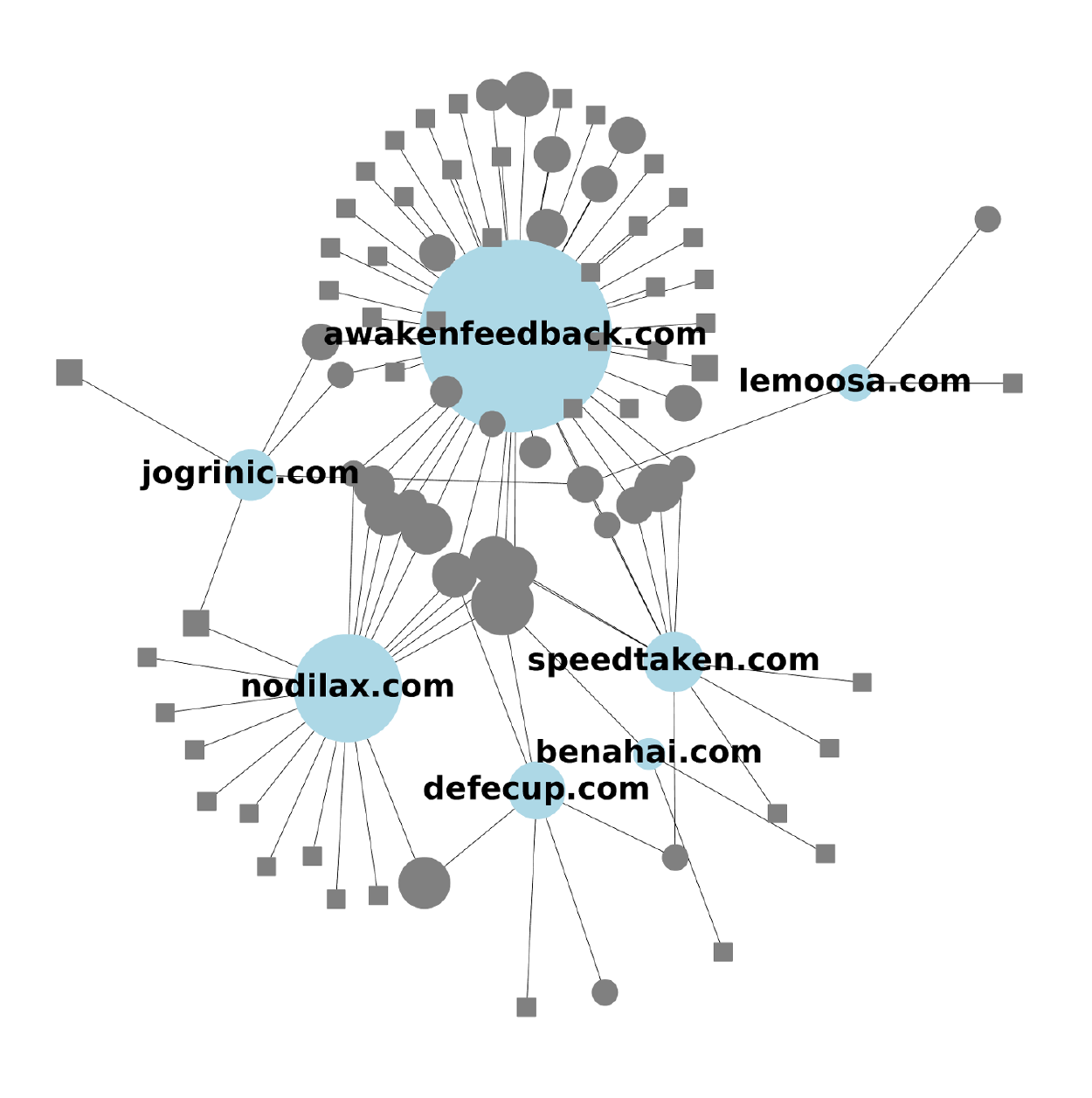}
        \setlength\abovecaptionskip{0.2\baselineskip}
        \caption{Meta Cluster of WPN-C2 WPNAC}
        \label{fig:suspicious_ads_2}
    \end{subfigure}%
    \begin{subfigure}{0.2\textwidth}
        \centering
        \includegraphics[scale=0.5]{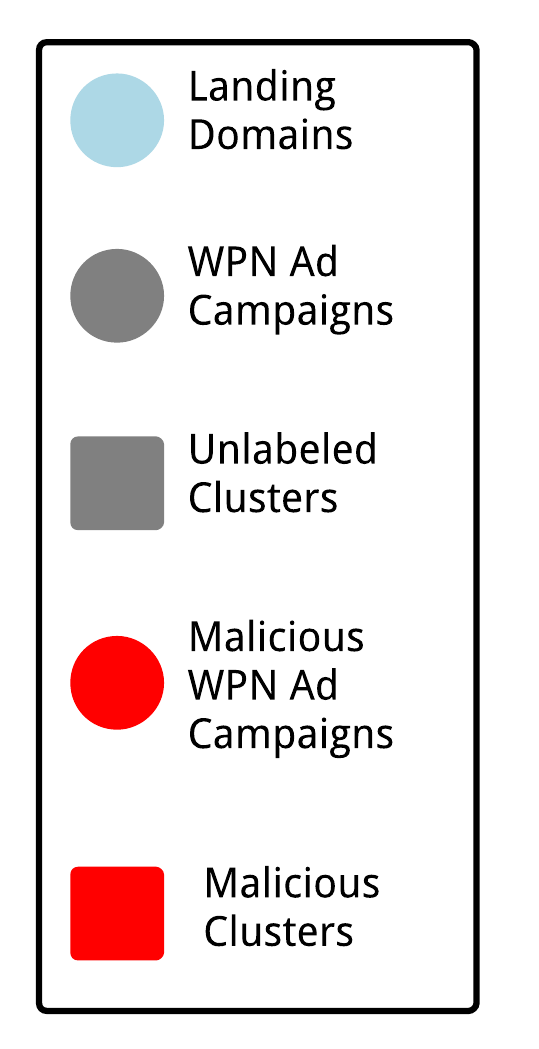}
    \end{subfigure}
    \setlength\abovecaptionskip{0.4\baselineskip}
    \setlength\belowcaptionskip{-0.5\baselineskip}
    \caption{Graphical Representation Examples of Meta Clusters}
    \label{fig:bi-partite}
\end{figure*}

Figure~\ref{fig:bi-partite} provides two examples of meta clusters.
Figure~\ref{fig:suspicious_ads_1} contains as a node cluster {\tt WPN-C1} from
Figure~\ref{fig:wpn_clusters}, as well as other 6 related WPN ad campaigns that
are likely orchestrated by the same operators. This meta-cluster contains many
{\em known malicious} WPN messages, and we therefore label it as suspicious. By
manual inspection, we verified that all domains involved in this meta cluster
host visually similar malicious pages (e.g., online survey scam pages).

Figure~\ref{fig:suspicious_ads_2} shows another example of meta-cluster, which
includes cluster {\tt WPN-C2} from Figure~\ref{fig:wpn_clusters} as a node,
along with {\em 30} other related WPN ad clusters. In this meta-cluster, none of
the WPN clusters (i.e., the nodes) were initially labeled as malicious by either
VT or GSB. However, we manually inspected all landing pages pointed to by WPN
messages including in the meta cluster, and we were able to confirm that these
are indeed malicious, in that they display fake PayPal messages and alerts that
lead users to survey scams and likely phishing-related pages.

\end{appendices}